\documentclass[a4paper]{jpconf}
\usepackage{graphicx}
\usepackage{placeins}
\usepackage{wasysym}

\newcommand{\hi}{heavy-ion\ }

\newcommand{\NICA}{~{NICA}}
\newcommand{\MPD}{~{MPD}}
\newcommand{\CF}{{CF}}
\newcommand{\HP}{{HP}}
\newcommand{\FHCal}{~{FHCal}}
\newcommand{\UrQMD}{~{UrQMD}}
\newcommand{\LAQGSM}{~{LAQGSM}}
\newcommand{\GEANT}{~{GEANT4}}

\long\def\/*#1*/{}

\begin{document}
\title{Performance of the MPD experiment for the anisotropic flow measurement}

\author{{P.~Parfenov$^{1,2}$}, {I.~Selyuzhenkov$^{1,3}$}, {A.~Taranenko$^1$} and {A.~Truttse$^1$}}

\address{$^1$ National Research Nuclear University MEPhI (Moscow Engineering Physics Institute), Kashirskoe highway 31, Moscow, 115409, Russia\\ $^2$ Institute for Nuclear Research of the Russian Academy of Sciences, Moscow, Russia\\ $^3$ GSI Helmholtzzentrum f{\"u}r Schwerionenforschung, Darmstadt, Germany}

\ead{petr.parfenov@cern.ch}

\begin{abstract}
The main goal of the future MPD experiment at NICA is to explore the QCD phase diagram in the region of highly compressed and hot baryonic matter in the energy range corresponding to the highest chemical potential. Properties of such dense matter can be studied using azimuthal anisotropy which is categorized by the Fourier coefficients of the azimuthal distribution decomposition. Performance of the detector response based on simulations with realistic reconstruction procedure is presented for centrality determination, reaction plane estimation, directed and elliptic flow coefficients.
\end{abstract}

\section{Introduction}
Studies of the quark-gluon matter thermodynamical properties is one of the main priorities in the number of experiments specializing in the \hi physics \cite{Czopowicz:2012ey}. Transverse azimuthally anisotropic flow measurements are one of the key methods to study the time evolution of the strongly interacted medium formed in the nucleous collisions. In the non-central collisions, initial spatial anisotropy results in the azimuthally anisotropic particle emission. The magnitude of the anisotropic flow is defined using the the Fourier coefficients $v_k\{\Psi_n\}$ of azimuthal distribution of the emitted particles with respect to the reaction plane \cite{Poskanzer:1998yz}:
\begin{eqnarray}
\frac{dN}{d(\varphi - \Psi_{n})} = 1+2\sum\limits_{k=1}^{\infty}v_{k}\cos \left[ k(\varphi - \Psi_{n}) \right],
\end{eqnarray}
where $\varphi$ -- is the azimuthal angle of particle, $k$ -- is the harmonic order and $\Psi_n$ is the $n$-th order symmetry plane angle. $v_1$ is hence called directed flow, $v_2$ -- elliptic flow.

In this  work centrality determination based on the multiplicity from TPC and anisotropic flow analysis for $Au+Au$ collisions will be presented for the two energies corresponding the highest and lowest eneries of the \NICA\ collider.

\FloatBarrier
\section{Simulation and analysis}
The future \MPD\ detector will be capable of a 4$\pi$-spectrometer, detecting charged hadrons,
electrons and photons in heavy ion collisions at high luminosities in \NICA\ energy range \cite{MPD:2017}. In
order to achieve this goal the detector comprises precise tracking system and highly-effective
particle identification system based on time-of-flight measurement and calorimetry.

Primary track selection based on the DCA distributions and implementation of the realistic tracking algorithm Cluster Finder (\CF) will be shown compared to the previous results \cite{Svintsov:2017rac}.

For the event generation the \UrQMD\ (Ultra-relativistic Quantum Molecular Dynamics) \cite{UrQMD} and \LAQGSM\ (The Los Alamos version of the Quark-Gluon String Model) \cite{LAQGSM} were used. The \UrQMD\ is used for performance study of the reaction plane determination and anisotropic flow measurements for the beam energy $5$ and $11\ GeV$ while the \LAQGSM is used for the reaction plane determination for the $11\ GeV$ only. Used statistics is $1$M events for each of the energy point and model choice. Further simulations were carried out using \GEANT\
framework using \MPD\ detector geometry and Cluster Finder (\CF) and Hit Producer (\HP) tracking algorithms for $5$ and $11\ GeV$ correspondingly.
Following cuts were used in the analysis:
\begin{itemize}
	\item $|\eta|<1.5$
	\item $0.2<p_T<3\ GeV/c$
	\item $N_{hits}^{TPC}>32$
	\item $ 2 \sigma $ DCA cut for primary particle selection
	\item particle identification (PID) -- is the cuts from PDG codes (Monte Carlo information)	 
\end{itemize}
where DCA -- is the distance of the closest approach between the reconstructed vertex and a charged particle track.

For the collective flow measurement event plane method was used \cite{Poskanzer:1998yz}. Reaction plane was estimated from the energy deposition of the nuclear fragments in backward and forward rapidities in the forward hadron calorimeters (\FHCal). $Q$-vector was calculated as follows:

\begin{eqnarray}
q_x^{m}=\frac{\sum E_i \cos m\varphi_i}{\sum E_i},\ q_y^{m}=\frac{\sum E_i \sin m\varphi_i}{\sum E_i}.
\end{eqnarray}
The event plane angle was calculated as follows:
\begin{eqnarray}
\Psi_m^{EP}=\textrm{TMath::ATan2}(q_y^m,q_x^m),
\end{eqnarray}
where $E_i$ is the energy deposition in the $i$-th module of \FHCal, $\varphi_i$ -- its azimuthal angle. For $m = 1$ weights had opposite signs for backward and forward rapidities due to the antisymmetry of the $v_1$ as a function of rapidity.
The values of $v_n$ itself could be calculated as follows:
\begin{eqnarray}
v_n=\frac{\left\langle\cos\left[n(\varphi - \Psi_m^{EP})\right]\right\rangle}{Res_n\{\Psi_m^{EP}\}},\ Res_n\{\Psi_m^{EP}\}=\left\langle\cos\left[n(\Psi_m^{EP}-\Psi_m)\right]\right\rangle,
\end{eqnarray}
where $Res_n\{\Psi_m^{EP}\}$ is the event plane resolution, $\Psi_m$ is the $n$-th order collision symmetry plane, which cannot be measured experimentally.
So, in order to estimate event plane resolution, the two-subevent method with extrapolation
algorithm was used \cite{Poskanzer:1998yz}:
\begin{eqnarray}
Res^2_n\{\Psi^{EP,A}_m,\Psi^{EP,B}_m\} = \left\langle\cos\left[n(\Psi_m^{EP,A}-\Psi_m^{EP,B})\right]\right\rangle,\ Res_n\{ \Psi^{EP}_m \} = Res_n(\sqrt{2}\chi_{A,B}),
\end{eqnarray}
where $\chi_{A,B}$ -- is the parameter proportional to the $v_n$ and $\sum E_i$, $A$ and $B$ represent two subevents - left(backward rapidity) and right(forward rapidity)
\FHCal\ detectors.
In this work $v_1$ and $v_2$ was measured with respect to $1$-st order event plane ($m=1$).

\FloatBarrier
\section{Results and conclusions}
\FloatBarrier
\subsection{Centrality determination}
\begin{figure}[h]
	\begin{minipage}{18pc}
		\includegraphics[width=18pc]{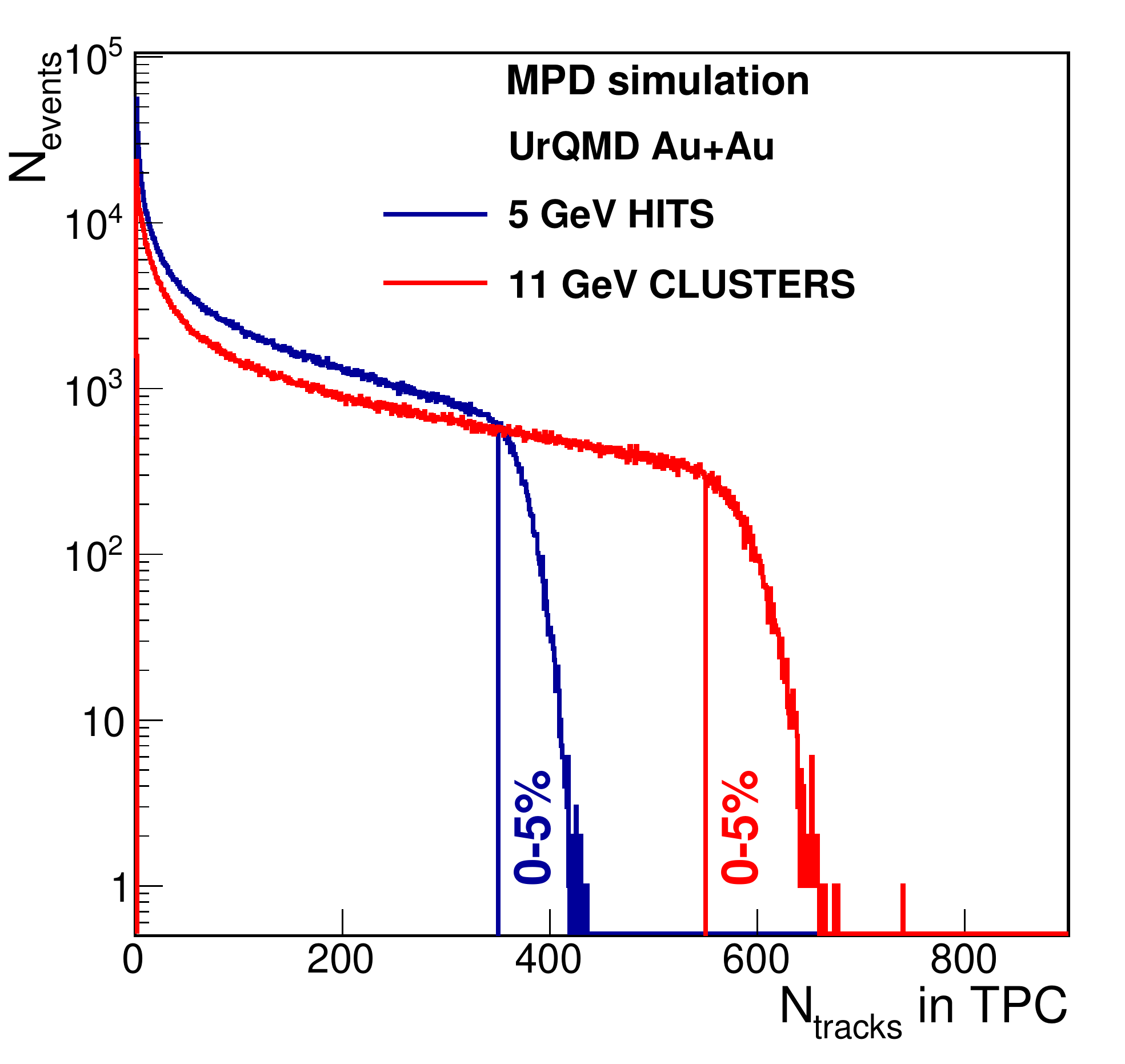}
	\end{minipage}\hspace{2pc}%
	\begin{minipage}{18pc}
		\includegraphics[width=18pc]{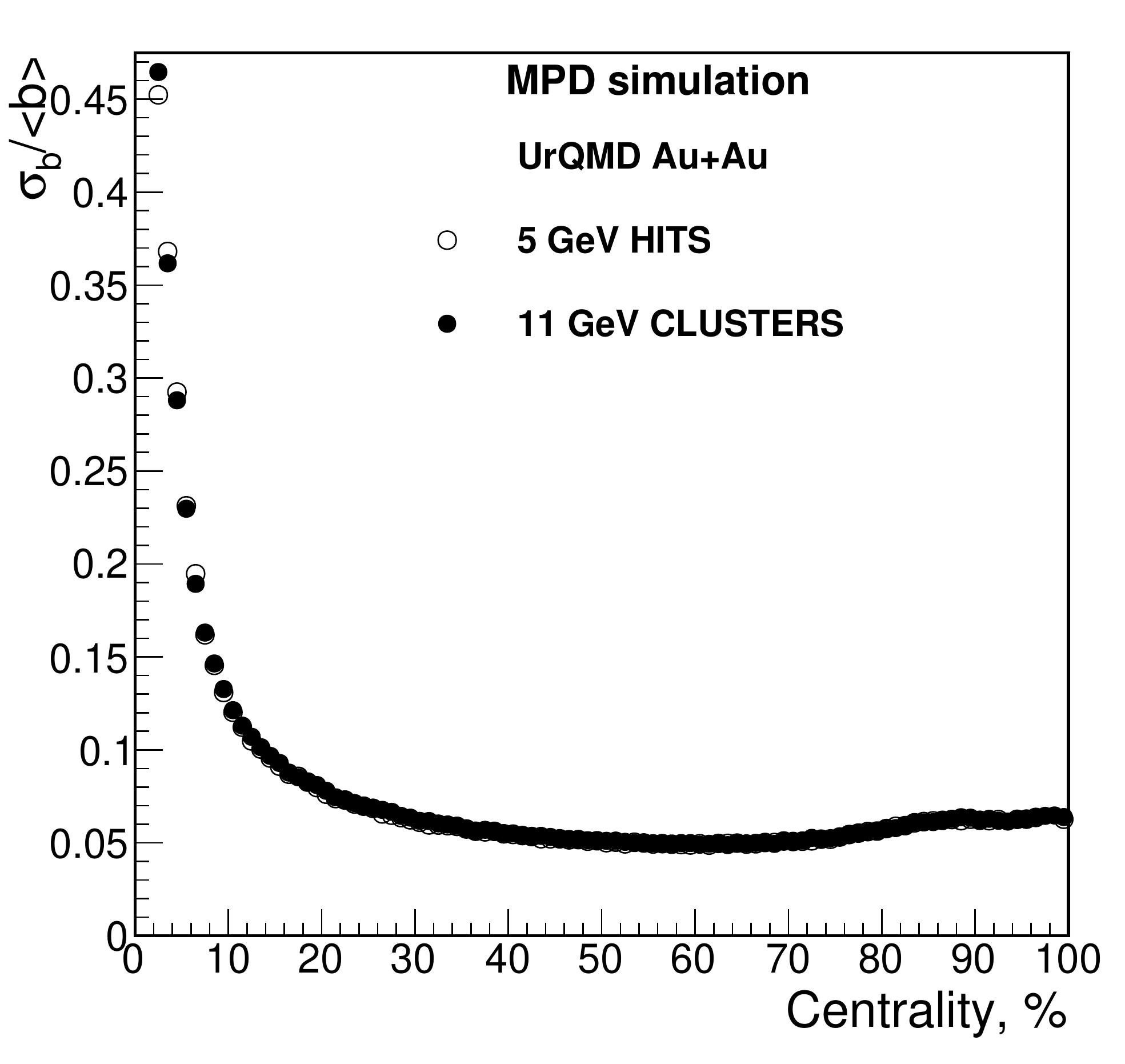}
	\end{minipage} 
\caption{\label{Mult} (left) Multiplicity distribution of the produced particles in TPC for $5$ and $11\ GeV$. Vertical lines indicate $0-5\%$ centrality range. (right) Relative width $\frac{\sigma_b}{\langle b \rangle}$ of the impact parameter $b$ distribution in the given centrality classes.}
\end{figure}

On the Figure \ref{Mult} (left) multiplicity of the primary charged particles produced in the $Au+Au$ collisions calculated using TPC detector is shown. This distribution was used to introduce the centrality classes with equal number of particles in each class. Centrality resolution of used classification is shown on the Figure \ref{Mult} (right). In the $10-80\%$ centrality range resolution $\frac{\sigma_b}{\langle b \rangle} \sim 5-10\%$ for both \CF\ and \HP\ tracking algorithms.
\FloatBarrier
\subsection{Azimuthal anisotropic flow}
\begin{figure}[h]
	\begin{minipage}{18pc}
		\includegraphics[width=18pc]{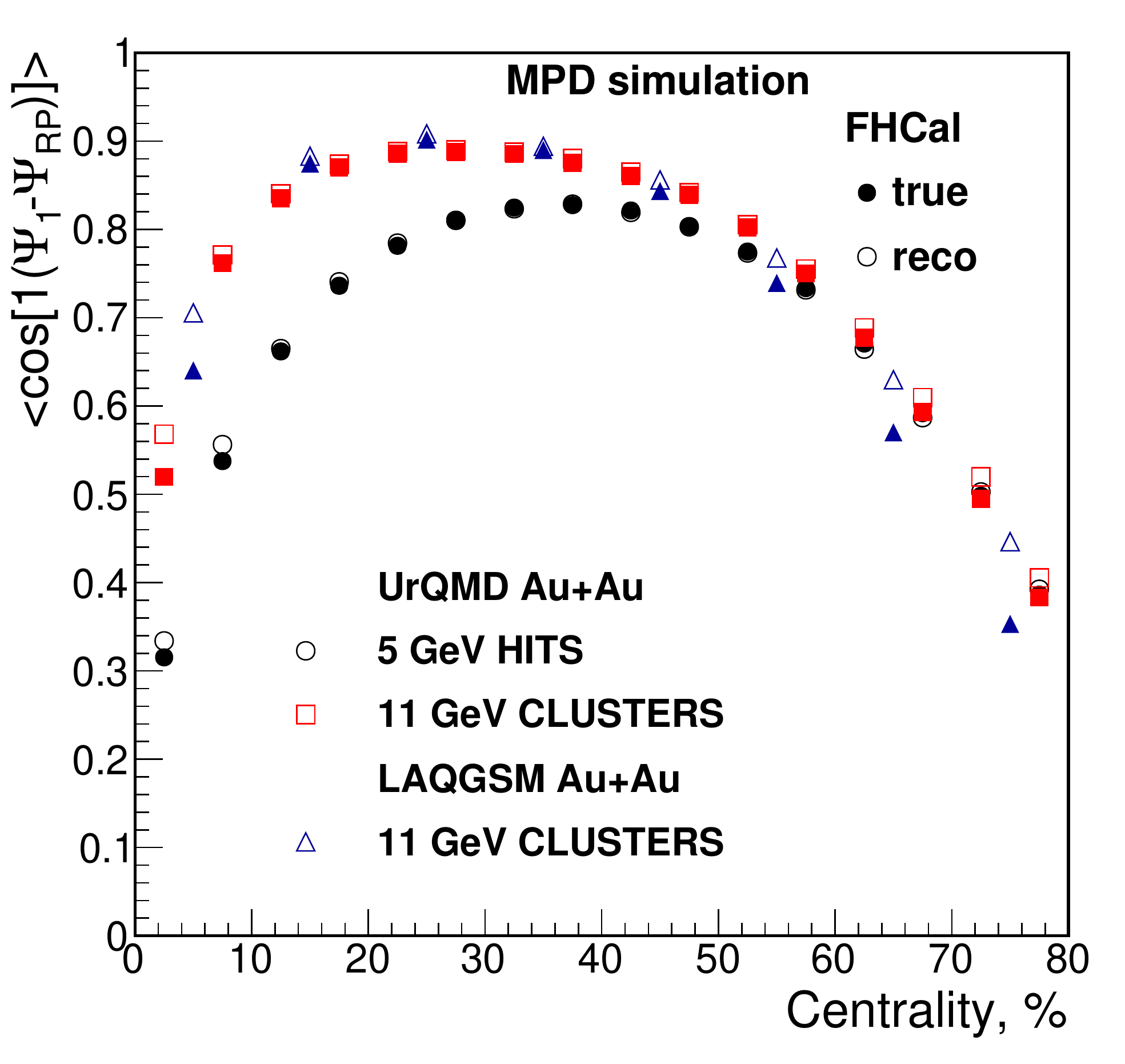}
	\end{minipage}\hspace{2pc}%
	\begin{minipage}{18pc}
		\includegraphics[width=18pc]{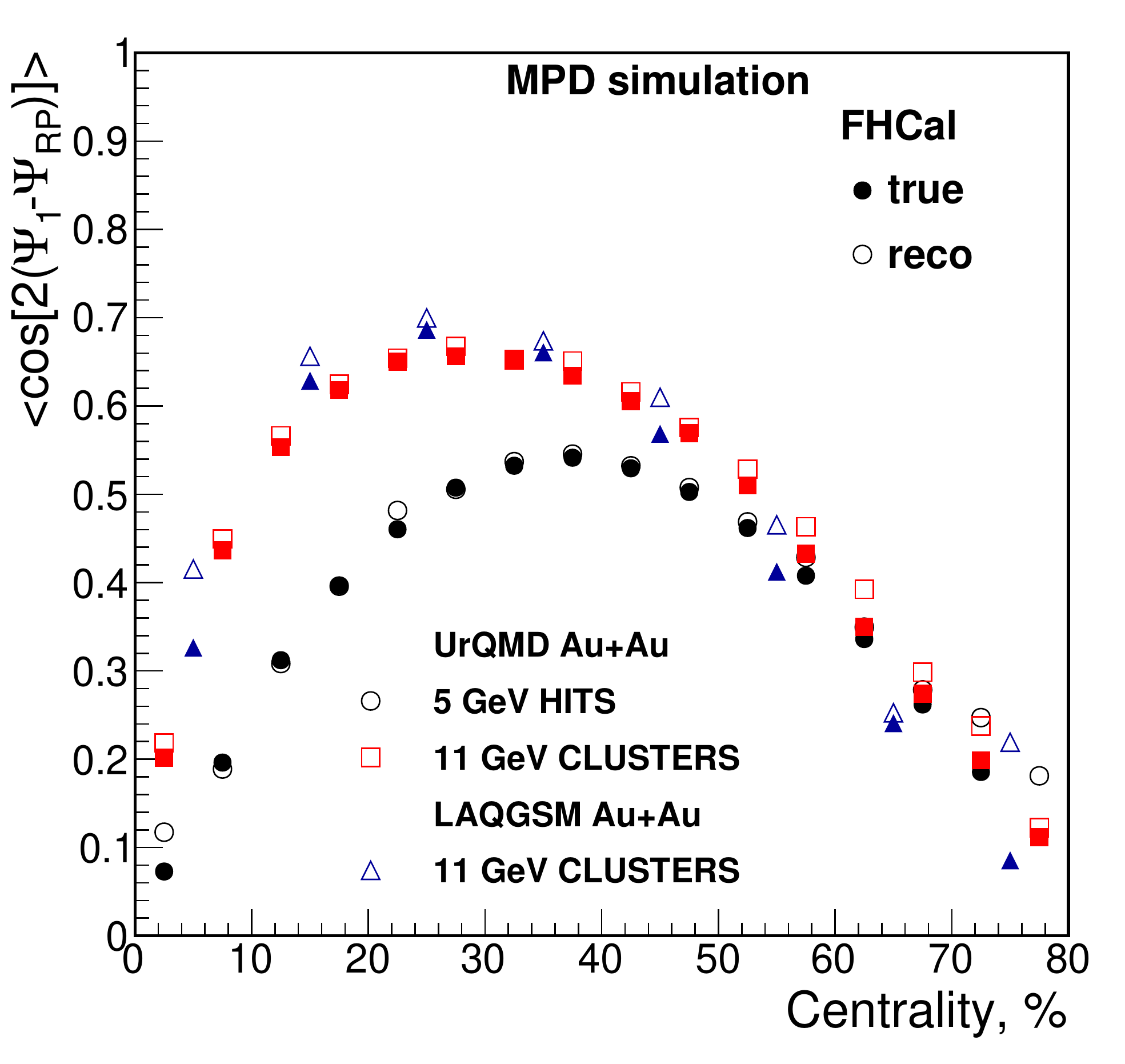}
	\end{minipage} 
	\caption{\label{Resolution}Resolution correction factor as a function of centrality for $v_1$ (left) and $v_2$ (right) for the \UrQMD\ and \LAQGSM\ event generators. Results from the \GEANT\ simulation marked as true and one from the reconstruction procedure is marked as reco.}
\end{figure}

On the Figure \ref{Resolution} resolution correction factor for $v_1$ and $v_2$ is shown. Since the \LAQGSM\ simulates nuclear fragments, one can see the deterioration of the resolution factor because more particles goes through the beam hole in the center of \FHCal\ unregistered. Other than that, results shows good performance in the wide centrality range $0-80\%$ for all energies and tracking algorithms.

\begin{figure}[h]
	\begin{minipage}{18pc}
		\includegraphics[width=18pc]{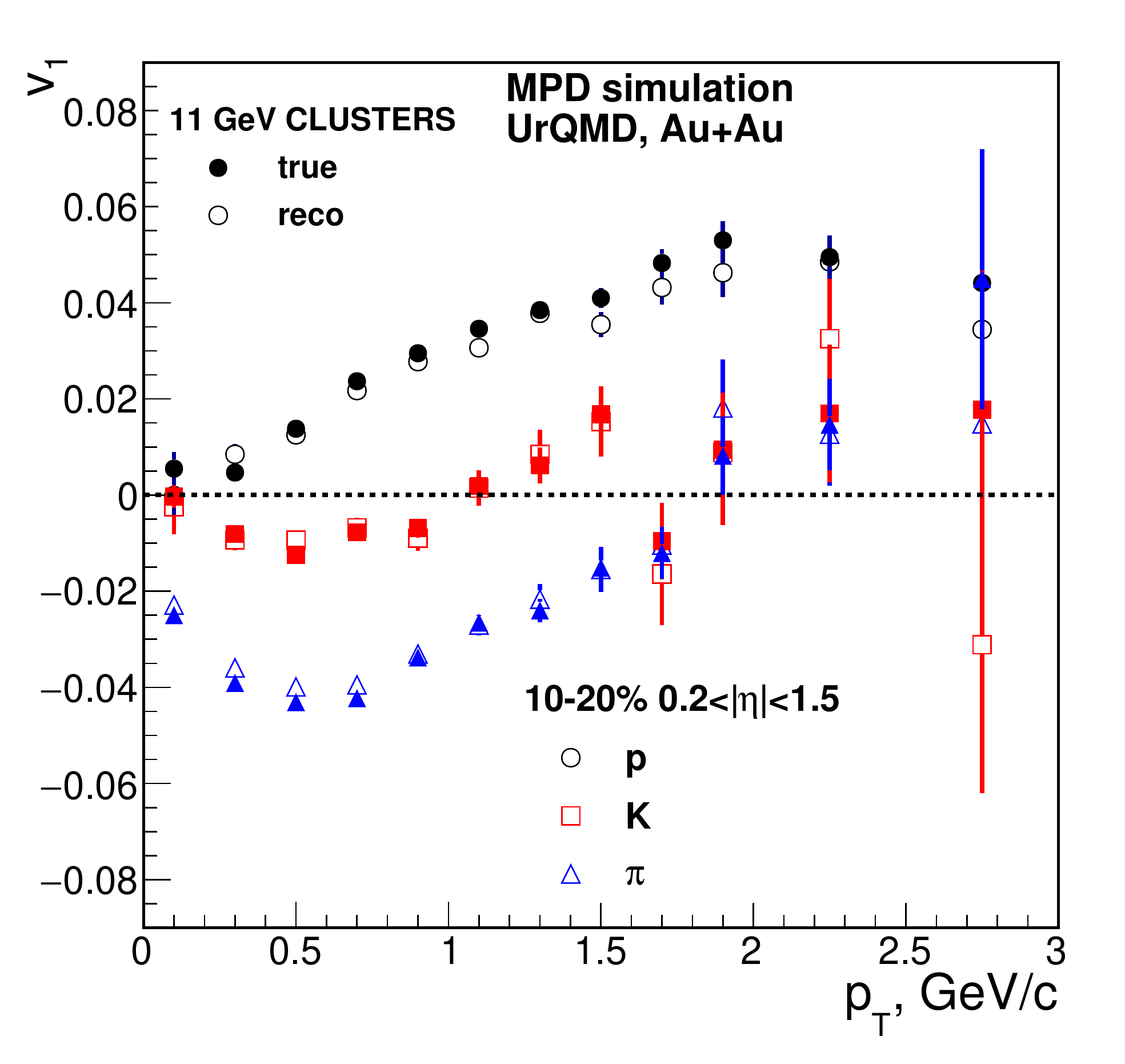}
	\end{minipage}\hspace{2pc}%
	\begin{minipage}{18pc}
		\vspace{0pc}\includegraphics[width=18pc]{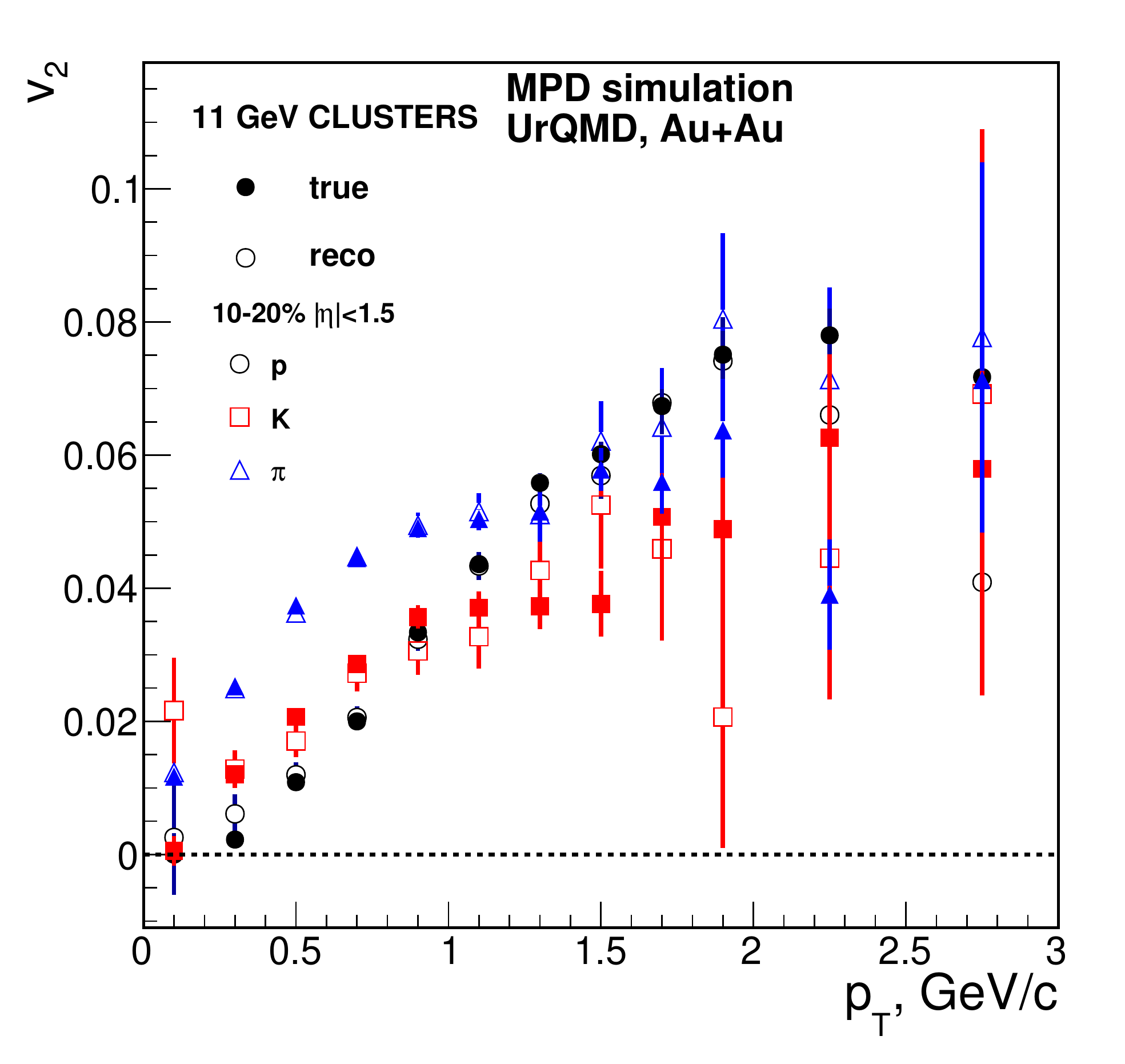}
	\end{minipage} 
	\caption{\label{vn}Directed flow $v_1$ (left) and elliptic flow (right) as a function of $p_T$. Signal from the \GEANT\ simulation marked as true and one from the reconstruction procedure is marked as reco.}
\end{figure}

On the Figure \ref{vn} one can see the directed $v_1$ and elliptic $v_2$ flow as a function of $p_T$. Signal after \GEANT\ simulation (true) is compared with one after reconstruction procedure (reco) which is how future experimental data will be analyzed. One can see that the difference between true and reco values is negligible.

\FloatBarrier
\section{Summary}

Track multiplicity of the emitted charged particles in TPC can be used for centrality determination with resolution $5-10\%$ in a wide centrality range $10-80\%$.
Event plane orientation can be estimated using energy deposition in \FHCal\ with high resolution factor ($Res_1\{\Psi^{EP}_1\}\sim 0.9,\ Res_2\{\Psi^{EP}_1\}\sim 0.7$ for centrality $20-40\%$).
Directed ($v_1$) and elliptic ($v_2$) flow were extracted in simulations using event plane method. Results for the reconstructed (reco) and generated (true) values are in good agreement.

\FloatBarrier
\section{Acknowledgments}
This work was partially supported by the Ministry of Science and Education of the Russian Federation, grant N 3.3380.2017/4.6, and by the National Research Nuclear University MEPhI in the framework of the Russian Academic Excellence Project (contract No. 02.a03.21.0005, 27.08.2013).

\end{document}